\documentclass[prb,twocolumn,preprintnumbers,amsmath,amssymb,floatfix,reprint]{revtex4}
\usepackage{hyperref}
\usepackage{epsfig}
\begin{document}

\title{Role of interface coupling inhomogeneity in domain evolution in exchange bias}

\author{Andrea~Benassi$^1$, Miguel A.~Marioni$^{1}$, Daniele Passerone$^1$, and Hans J.~Hug$^{1,2}$}

\affiliation{
$1-$ Empa, Swiss Federal Laboratories for Materials Science and Technology, CH-8600 D\"{u}bendorf, Switzerland.\\
$2-$ Department of Physics, Universit\"{a}t Basel, CH-4056 Basel, Switzerland.
}

\begin{abstract}
Models of exchange-bias in thin films have been able to describe various aspects of this technologically relevant effect.
Through appropriate choices of free parameters the modelled hysteresis loops adequately match experiment, and typical domain structures can be simulated.
However, the use of these parameters, notably the coupling strength between the systems' ferromagnetic (F) and antiferromagnetic (AF) layers, obscures conclusions about their influence on the magnetization reversal processes.
Here we develop a 2D phase-field model of the magnetization process in exchange-biased $\mathrm{CoO/(Co/Pt)}_{\times n}$ that incorporates the 10\,nm-resolved measured local biasing characteristics of the antiferromagnet.
Just three interrelated parameters set to measured physical quantities of the ferromagnet and the measured density of uncompensated spins thus suffice to match the experiment in microscopic and macroscopic detail.
We use the model to study changes in bias and coercivity caused by different distributions of pinned uncompensated spins of the antiferromagnet, in application-relevant situations where domain wall motion dominates the ferromagnetic reversal.
We show the excess coercivity can arise solely from inhomogeneity in the density of biasing- and anti-biasing pinned uncompensated spins in the antiferromagnet.
Counter to conventional wisdom, irreversible processes in the latter are not essential.
\end{abstract}

\maketitle

In coupled ferromagnetic- (``F'') and antiferromagnetic (``AF'') thin films exchange-bias can arise for fixed AF magnetic structures\cite{Ungurenau-2010a}.
The effect, widely used in contemporary magnetic devices such as giant- and tunnel-magnetoresistive thin-film sensors\cite{Baibich-1988a,Binasch-1989a}, is set up when the F structure ``imprints'' a stabilizing structure in the AF upon cooling below the AF N\'{e}el temperature.
It is manifested primarily as a lateral shift of the hysteresis loop of the F layer\cite{Meiklejohn-1956a}.
Often the width of the loop also increases with the onset of exchange-bias (e.g.~\cite{Qian-1998a,Fulcomer-1972a,Ali-2008a,Nogues-1999a}), recently prompting studies of the use of this excess coercivity as a proxy for the degree of sub-monolayer Co-oxidation\cite{Kosub-2012a}.\\
Microscopic model views of AFs and F-AF interfaces have provided insight into the mechanisms by which these features arise at the smallest scales\cite{Malozemoff-1987a,Mauri-1987a,Koon-1997a,Takano-1997a,Schulthess-1999a,Miltenyi-2000a,Nowak-2002a,Billoni-2011a}.
It is clear that one characteristic of the materials' systems associated with exchange bias is the existence of pinned uncompensated spins antiparallel to the F-magnetization\cite{Roy-2005a,Blackburn-2008a,Abrudan-2008a,Schmid-2010a} for Co, Fe or permalloy coupled to CoO.
The excess coercivity, on the other hand, has been circumscribed in the models to the effects of irreversible processes in the AF\cite{Nowak-2002a}.
Understanding these phenomena at scales that reveal domain wall motion in polycrystalline films is important because many (if not most) applications rely on magnetization reversal through this process -- or their impediment\cite{Stiles-1999a,Kappenberger-2003a,Schmid-2010a,Tieg-2010a}.
Hence the relevance of work by Fujiwara et al.~and Stiles et al.\cite{Fujiwara-1996a,Stiles-1999a,Stiles-2001a} that studied the influence of distributions of AF crystallite orientations and anisotropy in exchange bias.
However, in these studies the evolution of domain walls in the F during reversal could not be accounted for because of the exceedingly high computational cost of modelling macroscopic systems.
Other works have recently overcome this limitation\cite{Suess-2003a,Saha-2003a,Dorfbauer-2005a,Harres-2012a}.
But the role of the AF/F coupling, which is perhaps the least well understood component of exchange bias, is obscured by the reliance on free parameters for its description, and by the large variety of experimental results.

With this work our goal is to {\em avoid} free parameters in a model description of F reversal processes in typical exchange bias systems.
Instead, we want to rely exclusively on measured (or literature) values for the samples described, and show how accurate a model description can be, as assessed from measured microscopic domain images in applied fields and from hysteresis loops.
Accordingly, the dispersion in the measured sample data used as model input must be small -- hence we studied a single sample.
In particular, on the contentious issue of the distribution of pinned uncompensated spins, we can use 10\,nm-resolved experimental data from Schmid et al.\cite{Schmid-2010a}.
As for the coupling, its average can be deduced from magnetometry of the exchange-bias field $H_{ex}$.
Various statistics of the anisotropy of the AF on a granular scale can be accessed with techniques such as that proposed by Vallejo Fernandez et al.\cite{VallejoFernandez-2007a}.
This information would be essential for a correct account of temperature dependent- and training phenomena, and lacking it, we do not attempt to describe these effects (Incorporating them would be possible, but beyond the scope of this work).
The spatial distribution of F and AF anisotropy on the other hand cannot easily be furnished by experiment.
So as not to shape the model outcome with our particular choice of distribution, we select the most general distribution possible, a Gaussian one.
Improved models would base the anisotropy distributions on statistics of measured Barkhausen avalanches in the sample\cite{Benassi-2011a}, or infer spatial distributions from the marginal changes of domain boundaries with applied field at room temperature, where there is no exchange bias.

Our model is a 2D phase-field model, similar to those used to describe ferromagnetic films in relation to the role of disorder in domain dynamics.
For instance, return point memory effects~\cite{Pierce-2007a}, Barkhausen avalanche distributions~\cite{Benassi-2011a}, and the role of defects in the domain reorientation under the influence of an oscillating external field~\cite{Kudo-2009a} have been investigated in this fashion.
These studies did not attempt to match experiment quantitatively, in part because the local domain pinning strength was not known; nor have they been implemented in the context of exchange-bias.
From this perspective, our model differs from conventional ones in three important ways:
First, it describes an exchange-bias system on a scale relevant for magnetization reversal processes governed by domain wall motion.
Second, it incorporates an experimentally determined\cite{Schmid-2010a} 10\,nm-resolved distribution of F-AF pinned uncompensated spins ($^{pin}UCS$) over the $2\mu$m$\times 2\mu$m area modelled.
As we will show in the following, the inhomogeneity of pinned UCS affects $H_{ex}$ and $H_c$\cite{Takano-1997a,Stiles-2001a,Schmid-2010a}.
Hence the importance of using experimental values for pinned UCS, and the implied refinement over previous models.
And third, the model agrees quantitatively with the experimentally determined hysteresis loop and 10\,nm-resolved domain evolution, using for both scales one and the same set of material parameters that are in agreement with commonly accepted experimental values.

\section{Results}
\textbf{Model construction.}
\begin{figure}
\centering
\includegraphics[width=8.5cm,angle=0]{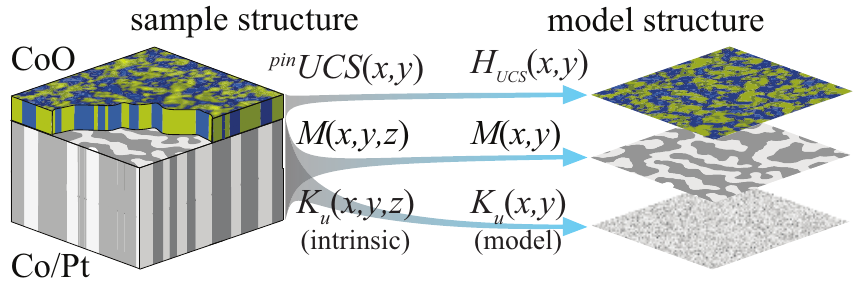}
\caption{Schematic of our exchange-biased $\rm{CoO_{1nm}/Co_{0.6nm}[Pt_{0.7nm}/Co_{0.4nm}]}_{\times20}$ multilayer.}
\label{fig:SampleOverview}
\end{figure}
Figure 1 illustrates schematically our $\rm{CoO_{1nm}/Co_{0.6nm}[Pt_{0.7nm}/Co_{0.4nm}]}_{\times20}$ exchange-biased system and the elements of the model counterpart.
In this work, we obtain the domain dynamics from the Landau-Lifshitz-Gilbert equation (LLG)\cite{Gilbert-2004a,Brown-1963a,Benassi-2013a} governing the damped precession of the magnetization $\mathbf{M}$ of the ferromagnetic layer in presence of a field $\mathbf{B}$:
\begin{equation}
\frac{\partial \mathbf{m}}{\partial t}=-\frac{\gamma}{1+\xi^2}\;\mathbf{m}\times\bigg[\mathbf{B}+  \xi \bigg( \mathbf{m}\times\mathbf{B}\bigg) \bigg],
\label{llg}
\end{equation}
where $\gamma$ is the gyromagnetic ratio, $\xi$ the Gilbert damping parameter, and $\mathbf{m}(\mathbf{R},t)=\mathbf{M}(\mathbf{R})/M_s$.
$M_s$ is the saturation magnetization, assumed uniform.
$\mathbf{B}$ is the magnetic field.
We write $\mathbf{B}=-1/M_s\,\delta\mathcal{H[\mathbf{m}]}/\delta \mathbf{m}+\mathbf{Q}(\mathbf{R},t)$, using the functional derivative of the hamiltonian $\mathcal{H}$ and a gaussian stochastic process $\mathbf{Q}(\mathbf{R},t)$ which accounts for the finite temperature effects
($\langle\mathbf{Q}(\mathbf{R},t)\rangle =0$ and $\langle \mathbf{Q}(\mathbf{R},t) \mathbf{Q}(\mathbf{R}',t')\rangle=\delta(t-t')\delta(\mathbf{R}-\mathbf{R}') 2 k_B T \xi / M_s \gamma$; $k_B$ is the Boltzmann constant and $T$ the temperature).\\
We assume uniform and small thickness $d$ and perpendicular magnetization.
The latter is represented by a scalar dimensionless field $m(x,y,t)$ such that $\mathbf{m}(\mathbf{r},t)=m(x,y,t)\hat{z}$.
This assumption implies the description will be accurate as long as the domain wall is much narrower than the domain width~\cite{Jagla-2005a,Jagla-2004a,Benassi-2013a}.\\
%%%%%%%%%%%%%%%%%%%%%%%%%%%%%%%%%%%%%%%%%%%%%%%%%%%%%%%%%%%%%%%%%%%%%%%%%%%%%%%%%%%%%%%%%%%%%%%%%%%%%%%%%%%%%%%%%%%%%%%%%%%%%%%%%%%%%%%%%%%%%%%%%%%%%%%%%%%%%%%%%%%%%%%%%%%%%%%%%%%%%%%%%%%%%%%%%%%%%%%%%%%%%%%%%%%%%%%%%%%%%%%%%%%%%%%%%%%%%%%%%%%%%%%%%%%%%%%%%%%%%%%%%%%%%%%%%%%%%%%%%%%%%%%%%%%%%%%%%%%%%%%%%%%%%%%%
The system's hamiltonian can then be written as:
\begin{eqnarray}
\nonumber
\mathcal{H}&=\int d^3\mathbf{R}\bigg[ -K_u(\mathbf{R})\frac{m^2}{2}+\frac{A}{2}(\nabla_{\mathbf{R}}m)^2+\frac{\mu_0 M_s^2 d}{8\pi}\\
\times &\int d^2\mathbf{R}'\frac{m(\mathbf{R}')m(\mathbf{R})}{\vert \mathbf{R}-\mathbf{R}'\vert^3} - \mu_0 M_s m (H_{ext}-H_{UCS}(\mathbf{R})) \bigg].
\label{hamilt}
\end{eqnarray}
The first term in (\ref{hamilt}) represents the anisotropy energy given the uniaxial anisotropy $K_u(\mathbf{R})$.
The second term in $\mathcal{H}$ represents the exchange interaction in the F layer, described by the exchange stiffness $A$.
The third term represents the long-range, non-local, stray field ($\mu_0$ is the vacuum permeability) for small film thickness.
The fourth term contains two fields.
On the one hand there is an external uniform field $H_{ext}$ whose value is quasi-statically ramped up and down in time;
On the other, an effective field $H_{UCS}(x,y)=H_{eb}\rho_p(x,y)/\langle\rho_p\rangle$, describing the local biasing effect in terms of the macroscopic, measured exchange-bias field $H_{eb}$ and the measured\cite{Schmid-2010a} $^{pin}\mathrm{UCS}$ density $\rho_p(x,y)$ (cf.~Fig.~2(g)).
Simulations of the field-cooled hysteresis loops require using $H_{UCS}^{sat}(x,y)$ as described in Fig.~2(h) (see See Supplementary Information for details on its construction from $H_{UCS}(x,y)$).
Substituting (\ref{hamilt}) in the over-damped limit $\xi\gg 1$ of eq.~(\ref{llg}) and in the approximation of thin domain walls\cite{Jagla-2005a}, the equation for the F domain dynamics becomes:
\begin{eqnarray}
\nonumber
\frac{\partial m}{\partial \tau}&=(1-m^2)\bigg( \alpha(1-p(\mathbf{r}))\: m-\frac{1}{4\pi}\int d^2\mathbf{r}' \frac{m(\mathbf{r}')}{\vert \mathbf{r}-\mathbf{r}'\vert^3}\\
&+h_{ext}(t)- h_{UCS}(\mathbf{r}) + q(\mathbf{r},\tau)\bigg)+\beta \nabla^2_{\mathbf{r}}m,
\label{dimensionless}
\end{eqnarray}
where we have introduced explicitly the dimensionless units $\mathbf{r}=\mathbf{R}/d$ , $\tau=t \gamma \xi \mu_0 M_s$, $h_{ext}=H_{ext}/M_s$, $h_{UCS}=H_{UCS}/M_s$ and $q(\mathbf{r},\tau)=Q(\mathbf{R},t)/\mu_0 M_s$.
The dimensionless constants $\alpha=\langle K_u\rangle/\mu_0M_s^2$ and $\beta=A/\mu_0 M_s^2 d^2$ are the reduced anisotropy and exchange stiffness, respectively.
$p(\mathbf{r})$ is the distribution that accounts for the anisotropy inhomogeneity, and is not exactly known.
To avoid introducing overly restrictive assumptions into the model we choose $p(\mathbf{r})$ to be Gaussian-distributed spatially-uncorrelated noise with $\langle p \rangle=0$ and variance $\eta$, i.e.~$\langle p(\mathbf{r})p(\mathbf{r'})\rangle=\eta\delta(\mathbf{r}-\mathbf{r}')$.
Together, $\alpha$, $\beta$ and $\eta$ constitute the only parameters of the model.

\begin{figure}
\centering
\includegraphics{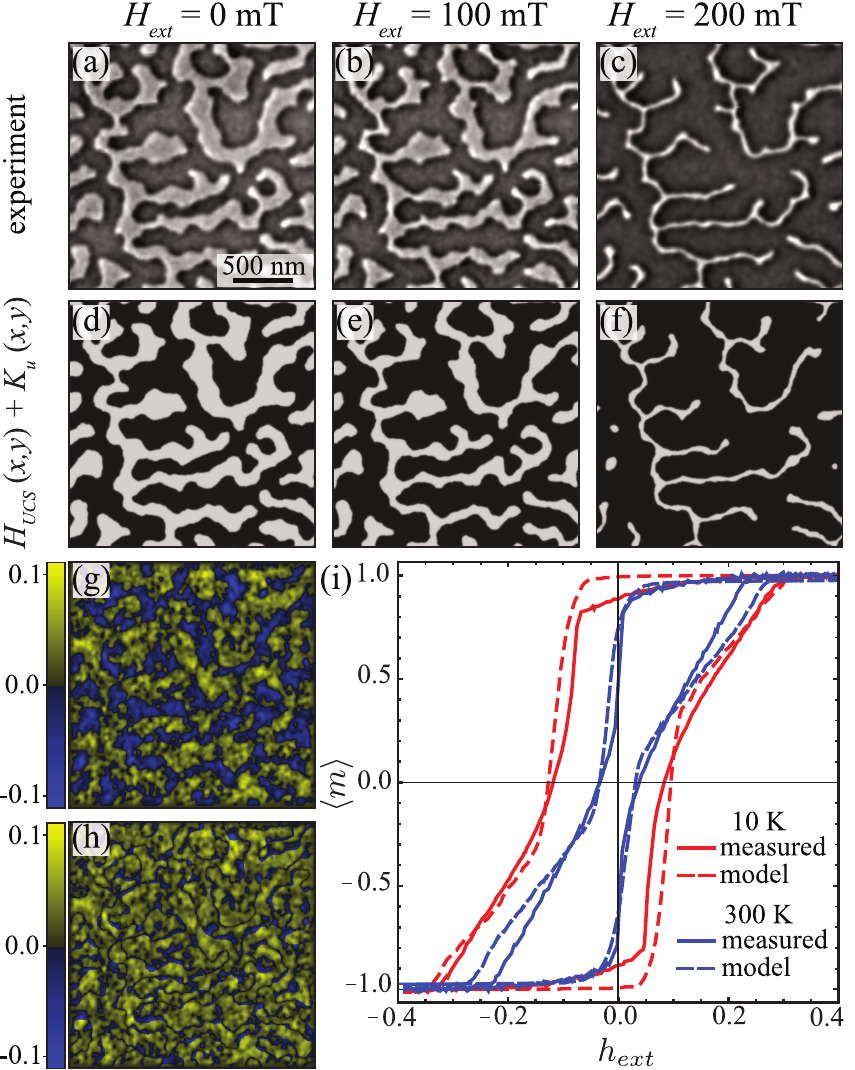}
\caption{
Model validation.
(a) Experimental 10\,K domain pattern at 0\,mT applied field (from Ref.~\cite{Schmid-2010a}).
(b) and (c) Corresponding domain patterns at 100\,mT and 200\,mT applied field.
(d), (e) and (f) Simulated domain patterns.
(g) Experimental pattern $h_{UCS}^{(exp)}$ used for (d) -- (e).
(h) Modified $h_{UCS}$ pattern ($h_{UCS}^{sat}$) for the simulation of hysteresis loops, obtained from (g) through inversion of the areas corresponding to the light domains of (d).
(i) Comparison of hysteresis loops from experiment\cite{Schmid-2010a} with the model result.
}
\label{fig:ModelValidation}
\end{figure}

\textbf{Model validation with experiment.}
In Fig.~2 we compare the model results for $\alpha=6.6$, $\beta=0.14$ and $\eta=1.88\times10^{-4}$ with experimental results from\cite{Schmid-2010a}.
Figures 2 (a) -- (c) are the experimental, 10\,K domain structures of the zero-field cooled system for 0\,mT, 100\,mT and 200\,mT applied field.
At 0\,mT Fig.~2 (d) is the stable domain structure calculated with the model starting with the domain structure from 2 (a).
The detailed resemblance between Fig.~2 (a) and (d) shows the measured pattern is stable for the model, given the magnetization dynamics it describes.
Starting from Fig.~2 (d) and gradually increasing the applied field, and letting the system evolve using Eq.~\ref{dimensionless} to a stable domain structure results in  Figs.~2 (e) and (f). The shape of the simulated domains matches the experimental one, and a high level of detail is reproduced by our model, although some discrepancies are apparent.
We then simulate a magnetometry measurement at 10\,K after cooling in a 1\,T field, that is, a typical exchange-bias measurement, Fig.~2 (i).
This we do using exactly the same values of $\alpha$, $\beta$ and $\eta$ employed for the above domain evolution.
However, we cannot use exactly the same distribution of pinned UCS as before, since it is set by the F's magnetization structure during cooling, which is now different.
Neither can we directly measure the pinned UCS from this experiment preparation\cite{Joshi-2011a}:
Recall that a uniformly magnetized F film would produce no stray field for the MFM to detect, so in the field cooled case the MFM would only image the inhomogeneity of the pinned UCS on a local scale, but not their average density.
But note that the F, over the areas inside its domains, sets the pinned UCS in the same way as an F saturated with the appropriate orientation would.
Thus for the loop simulation we construct the map of pinned UCS shown in Fig.\,\ref{fig:ModelValidation}\,(h)  from the one obtained after zero-field cooling, shown in  Fig.\,\ref{fig:ModelValidation}\,(g) by inverting the pUCS under the white ferromagnetic domains.
The important loop characteristics agree well with experiment.
Specifically, we observe a mean coercivity of 86.20\,mT and an exchange field of 12.35\,mT in the field-cooled case, which compare well with experiment (77.7\,mT and 13.6\,mT, respectively).
Prominent features of the magnetization loop, such as the knee at the nucleation field and the subsequent more protracted approach to saturation are also displayed by the model results for the hysteresis.
In the zero field cooled case, where we expect no macroscopic exchange bias, we obtain a match of comparable quality as for the field-cooled case.
For the corresponding comparison at 300\,K, also in Fig.~2 (i), we use $\alpha=5.8$ ($\beta$ and $\eta$ retain their 10\,K-values), thus accounting in a qualitative manner for the reduced anisotropy at higher temperatures.
Furthermore, consistent with the absence of $^{pin}\mathrm{UCS}$ above the AF blocking temperature, we set $h_{UCS}\equiv 0$ in this case.
As expected, the simulated loops are symmetric and have a considerably reduced coercive fields.
%--------------------------------------------
%--------------------------------------------

\textbf{Contributions to domain dynamics from anisotropy- and coupling-inhomogeneity.}
Our model allows us to investigate how the domains evolve greater detail, for example looking at intermediate field levels (See Supplementary Information for a video of the simulated F domain evolution and detailed images of domain patterns.).
Also, for instance, we can investigate the changes in domain evolution in the hypothetical case of $h_{UCS}\equiv 0$, that is, when domain boundary pinning is controlled solely by anisotropy inhomogeneity in the F layer and there is no net exchange-coupling between F and AF.
Figures 3(a) -- (c) show the resulting domain patterns at 0, 100, and 200\,mT in black and white.
For comparison, we superimpose in yellow the contours of the corresponding domain boundaries from experiment.
Likewise, we study the converse case of $h_{UCS}=h_{UCS}^{exp}$ and $\eta = 0$, i.e.~$K_u(\mathbf{R})=\langle K_u\rangle$, when pinning from anisotropy inhomogeneity is negligible.
We show the results in Figures 3(d) -- (f) (cf.~Supplementary Information at for a side-by-side summary of the images).
%--------------------------------------------
\begin{figure}
\centering
\includegraphics{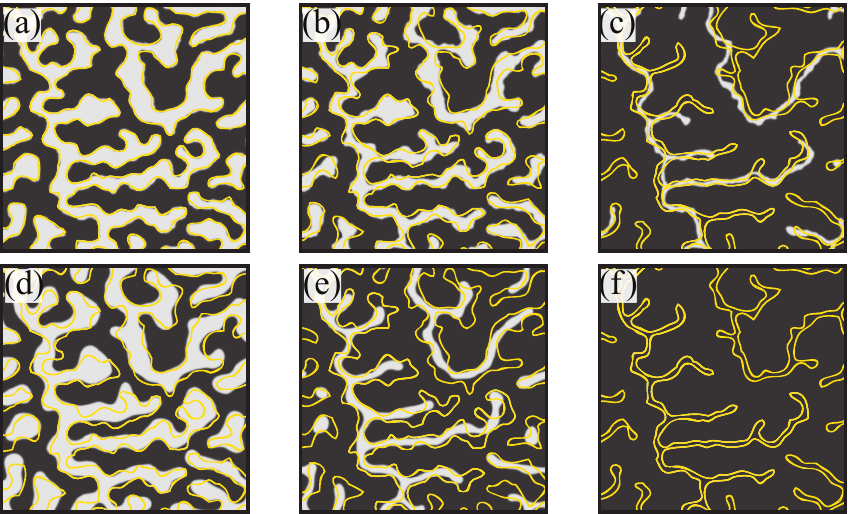}
\caption{
Simulated 10\,K magnetization patterns (black \& white) for different hypothetical model inputs, put in comparison with experiment (yellow trace).
(a) Simulated domains at 0\,mT applied field using $\eta=1.88\times10^{-4}$ and $h_{UCS} =0$.
(b) Idem  at 100\,mT.
(c) Ibidem at 200\,mT.
(d) Simulated domains at 0\,mT applied field using $\eta=0$ and $h_{UCS} =h_{UCS}^{exp}$ from Fig.~2(g).
(e) Idem  at 100\,mT.
(f) Ibidem at 200\,mT.
In this simulation the sample already saturates at this field level.
}
\label{fig:DomainSimulation}
\end{figure}
%--------------------------------------------

\textbf{Exchange-bias and coercivities for different inhomogeneity in local density of pinned uncompensated spins.}
With our model it is also straightforward to calculate the magnetization process characteristics for hypothetical distributions of pinned uncompensated spins $h_{UCS}$, i.e.~varying degrees of F-AF coupling inhomogeneity.
In Tab.~\ref{table1} we summarize the coercivity and exchange-bias fields from experiment and the full simulation, as well as values reported by simulations where $h_{UCS}$ has been modified arbitrarily.
\begin{table}
  \begin{tabular}{|l|c|c|}
    \hline
    {} & $\mu_0H_{eb}$ (mT) & $\mu_0H_c$ (mT) \\
    \hline
    \multicolumn{1}{|l|}{\bf{Experiment}}     & 13.6   & 77.7 \\
    \hline
    \multicolumn{3}{|l|}{ \bf{Model} $h_{UCS} = $} \\
    \hline
    \multicolumn{1}{|c|}{$h_{UCS}^{sat}$}                                         & 12.35   & 86.20 \\
    \multicolumn{1}{|c|}{0}                                                       & $<$ 0.01 & 84.93 \\
    \multicolumn{1}{|c|}{$h_{UCS}^{sat} + \langle h_{UCS}^{sat}\rangle$}          & 24.32   & 88.21 \\
    \multicolumn{1}{|c|}{$2\times\,h_{UCS}^{sat} - \langle h_{UCS}^{sat}\rangle$} & 12.02   & 92.32 \\
    \hline
  \end{tabular}
  \caption{Summary of the experimental and theoretical results from hysteresis loops at $T=10 K$.  }
  \label{table1}
\end{table}
The first two reference data sets comprise the simulation with $h_{UCS}=h_{UCS}^{sat}$ discussed in Fig.~2 (See also Supplementary Information), and a further simulation setting $h_{UCS}\equiv 0$, which confirms that the resulting loop is centered around 0\,mT.
It has a slightly reduced coercivity, and there is no exchange-bias in this case: $h_{eb}=0$.\\
Next, we can set $h_{UCS}=h_{UCS}^{sat}  + \langle\,h_{UCS}^{sat}\rangle$ to simulate a film with double the average density of $h_{UCS}^{sat}$ and with the same inhomogeneity.
In other words, a film with greater average $^{pin}\mathrm{UCS}$ but comparable levels of coupling frustration and local variations in the pinning ability of the AF.
The resulting hysteresis curve is laterally shifted by an amount twice as large as $H_{eb}$ in the first simulation, confirming the experimental results based on microscopic domain pattern evolution.
Notably the enlarged coercivity, which typically accompanies exchange-bias, remains at the original levels.\\
These findings change if, conversely, we set $h_{UCS}=2\times\,h_{UCS}^{sat} - \langle\,h_{UCS}^{sat}\rangle$ to simulate a film with the original average density of $^{pin}\mathrm{UCS}$ but twice the amplitude of the inhomogeneity.
In that case the model results in a greater coercivity without significant changes in the exchange-bias.

\section{Discussion}
From Figure 3 we can see that simulations without anisotropy inhomogeneity or without pinned uncompensated AF spins cannot match the experimental domains as accurately as the full simulation with $h_{UCS}=h_{UCS}^{exp}$ and $\eta=1.88\times10^{-4}$, Figs.~2(d)--(f).
Despite such deficiencies, the simulation with $h_{UCS}\equiv 0$ tracks the experiment with reasonable accuracy for 0 and 100\,mT applied field, and shows more prominent deviations from it only at 200\,mT.
This high field behavior is compatible with the smaller number of energy minima of sufficient strength in the absence of coupling to the AF.\\
As a means to control domain boundary motion $h_{UCS}$ appear to be slightly less effective than $K_u$ inhomogeneity, which agrees with $H_{ex} < H_c$.
In particular, Figs.~3(d) -- (f) depart markedly from the other simulations already at 0\,mT applied field.
Furthermore at 200\,mT the simulation would predict magnetization saturation, so, clearly, pinning from $h_{UCS}$ alone is unable to describe the measured domain structure at 200\,mT.
These observations do not imply a subordinate role of $^{pin}\mathrm{UCS}$ in exchange-bias; on the contrary.
They show that even without anisotropy inhomogeneity to pin the F domain walls a magnetization structure is retained up until at least 100\,mT, which would not be possible if $^{pin}\mathrm{UCS}$ did not pin the domains.\\
Moreover, the discrepancies between model and experiment depend significantly on the local values $K_u(\mathbf{R})$, the distribution of which we have carefully kept in the most generic form.
Because of that, $K_u(\mathbf{R})$ very likely differs from the real distribution of $K_u$, and simulation inaccuracies are to be expected.
Nevertheless, the model captures the essential physics of the magnetization reversal in the presence of domain wall motion.
An anisotropy distribution with a sufficiently large number of free parameters could of course be adjusted so as to yield a more perfect match between model and experiment.
However, this strategy would obscure, rather than clarify the magnetization reversal mechanism here.\\
We can gain additional insight into how the reversal is affected by the $^{pin}\mathrm{UCS}$ and $K_u$ distributions when we look at simulations of the magnetization loops.
For instance, we can reexamine the link between exchange-bias and average $^{pin}\mathrm{UCS}$ found in microscopic observations\cite{Schmid-2010a}.
Hysteresis loops at 10\,K are the macroscopic counterpart to these observations, which could not have been carried out in experiment due to the impossibility of arbitrarily changing the density of pinned uncompensated spins.\\
Table \ref{table1} confirms that exchange-bias is roughly proportional to the average density of pinned uncompensated spins.
In particular, pinned uncompensated AF spins aligned parallel to the F are detrimental to exchange bias.
Moreover, the inhomogeneity of the $^{pin}\mathrm{UCS}$ leads to an excess coercivity.
The importance of this finding is that rotating UCS and irreversible processes in the AF are not a necessary condition for excess coercivity, as the prevalent thinking holds\cite{Nowak-2002a}.
Nor is their role in F reversal at low temperature expected to be major, except perhaps in the case of strongly coupled\cite{Aley-2011a,Tsunoda-2006a,Tsunoda-2006b} F-AF layers.\\
The picture that emerges is one where the inhomogeneity of the F's anisotropy and of the AF's pinned uncompensated spins largely determines the details of the F reversal.
From our data\cite{Kappenberger-2003a,Kappenberger-2005b,Schmid-2008a,Schmid-2010a} as well as from XMCD measurements\cite{Blackburn-2008a} it is clear that the pinned uncompensated spins exist in exchange bias systems in relatively high areal densities, exceeding $\approx$10\% of a monolayer.
This is in apparent contradiction to the observed exchange fields, which are far smaller than would be expected if all pinned UCS coupled with bulk-order exchange constants to the F (note that by measuring {\em pinned} UCS we consider only the part that is stable in applied fields).
The problem stems from the default assumption that all pinned UCS participate in the coupling to the F.
Notice though, that the pinned UCS found in the aforementioned experiments align antiparallel to the F moments.
Consequently they cannot be aligned directly by the cooling field.
Instead, these pinned UCS must be exchange-coupled to the F (antiferromagnetically, possibly via a superexchange mechanism), or otherwise exchange-coupled to such UCS.
In either case it follows that these pinned UCS must be located at or near the F-AF interface.
But of them, only the ones that couple directly (antiferromagnetically, as discussed) to the F generate an exchange bias effect.
The remaining ones do not provide additional coupling, explaining the weakness of the exchange bias in the presence of a surprisingly high density of pinned UCS.
Our conclusion is supported by reflectometry experiments that revealed the pinned UCS existed over a film thickness larger than the roughness of the interface\cite{Blackburn-2008a}.
Hence models assuming a sharp interface or uncompensated spins located solely at an atomically sharp interface between an F and AF seem inappropriate to explain exchange bias.\\
Thus our 2D phase-field model of exchange-bias systems  based on general assumptions and parameters set by experiment, matches experiment in macroscopic and microscopic detail, and on this basis is able to establish that:
1) The magnitude of the average $^{pin}\mathrm{UCS}$ density determines the exchange-bias field $H_{eb}$ in spite of the fact that only a part of these $^{pin}\mathrm{UCS}$ couple directly to the F.
2) The spatial inhomogeneity of the $^{pin}\mathrm{UCS}$ governs the evolution of the domain pattern on a local scale and gives rise to excess coercivity associated with exchange bias.
3) Irreversible AF processes and $^{pin}\mathrm{UCS}$ rotation need not be invoked to explain excess coercivity.
4) The average coupling between the F moments and the $^{pin}\mathrm{UCS}$ is weak compared to intrinsic coupling constants describing the exchange in ferromagnets and antiferromagnets.
%%%%%%%%%%%%%%%%%%%%%%%%%%%%%%%%%%%%%%%%%%%%%%%%%%%%%%%%%%%%%%%%%%%%%%%%%%%%%%%%%%%%%%%%%%%%%%%%%%%%%%%%%%%%%%%%%%%%%%%%%%%%%%%%%%%%%%%%%%%%%%%%%%%%%%%%%%%%%%%%%%%%%%%%%%%%%%%%%%%%%%%%%%%%%%%%%%%

\section{Methods}
It is important to discuss the extent of the adjustments admitted in the course of matching the model results to experiment (See Supplementary Information for a table of material parameters and adjustment guidelines).
We emphasize that $\alpha$, $\beta$ and $\eta$ are not arbitrary (``free'').
Specifically, $M_s$ and $d$ are precisely known for the particular film we used\cite{Schmid-2010a}.
With regard to $\alpha$, because $\langle K_u\rangle$ is not available for our film the model calculations are carried out using a literature value\cite{Hubert-1998a} subject to fine adjustments of few \%.
This is acceptable given $K_u$ may differ slightly for two films of the same nominal characteristics and fabrication process.
We restrict the RMS inhomogeneity of $K_u$ to not more than about 20\%, effectively limiting the range of possible values of $\eta$ to small positive numbers of order $10^{-4}$.
Determining $\beta$ further requires specifying $A$.
We use bulk values found in literature\cite{Hubert-1998a} and apply corrections for the different dimensionality of our model, as called for by data on the domain wall width $\delta_{dw}$ in our system\cite{Hubert-1998a} and the relation $\delta_{dw}=\pi d\sqrt{\beta/\alpha}$ for Bloch-walls.
The values $\alpha_{ini}=6.25$, $\beta_{ini}=0.135$ and $\eta_{ini}=1.5\times10^{-4}$, used for the first model calculation on the basis of which further adjustments follow, are determined in this way.\\
The simulation of the domain pattern evolution over a series of applied fields consists in integrating Eq.~(\ref{dimensionless}) numerically (in Fourier space, to circumvent the non-locality. Cf.~Supplementary Information for technical details) starting from the known experimental\cite{Schmid-2010a} zero-applied-field F-domain structure (at $T=10K$), Fig.~2(a).
Hysteresis loops follow trivially from the series of patterns.
Taking into account the influence of $\alpha$, $\beta$ and $\eta$ on the domain evolution, we modify them slightly so that the zero applied field magnetization pattern in Fig.~2(a) becomes a stationary state of our model.
Even smaller manual adjustments of $\alpha$, $\beta$ and $\eta$ follow, to ensure the {\em simulated evolution} of the domain pattern with the applied field about matches the other available experimental domain patterns, i.e.~Fig.~2(b)--(c), at $h_{ext}=100$ and $200$\,mT, respectively.
A final manual fine-tuning of parameters seeks a fit of the experimental 10\,K hysteresis loop, Fig.~2(i).
Note that an automated search for best fitting $\{\alpha, \beta, \eta\}$ values is possible (e.g.~relying on cross-correlations for domain patterns as a fitness function\cite{Pierce-2007a}) but not practical at the moment.
It would not change the model conclusions substantially.
By this process we arrive at $\alpha=6.6$, $\beta=0.14$ and $\eta=1.88\times10^{-4}$, which depart only slightly from the initial values.\\

{\bf Acknowledgments -}  The experimental data supporting this work was previously published in I. Schmid et al. \emph{Phys. Rev. Lett.} \textbf{105} 197201 (2010), and comprises the measurements of I. Schmid, S. Romer and P. Kappenberger of a sample supplied by M.J. Carey, O. Hellwig and E.E. Fullerton, whom we would like to acknowledge especially. We would like to thank S. Zapperi and E. A. Jagla for helpful comments. This work has been supported by grant CRSII2 136287/1 from the Swiss National Science Foundation.\\

{\bf Author contributions -} 
M.A.M. and H.J.H. produced the main idea of the paper. A.B. developed and implemented the model with contributions of D.P., M.A.M. and H.J.H. All authors contributed to the scientific discussions, refining and clarifying the concepts presented by the manuscript, read and discussed the paper.\\

%%%%%%%%%%%%%%%%%%%%%%%%%%%%%%%%%%%%%%%%%%%%%%%%%%%%%%%%%%%%%%%%%%%%%%%%%%%%%%%%%%%%%%%%%%%%%%%%%%%%%%%%%%%%%%%%%%%%%%%%%%%%%%%%%%%%%%%%%%%%%%%%%%%%%%%%%%%%%%%%%%%%%%%%%%%%%%%%%%%%%%%%%%%%%%%%%%

\end{document}